%% file: hp.tex
\documentclass{LMCS}
\usepackage{enumerate}
\input{maquereau.sty}
\usepackage{hyperref}

\def\doi{4 (2:4) 2008}
\lmcsheading%
{\doi}
{1--20}
{}
{}
{Sep.~20, 2007}
{May.~14, 2008}
{}   

\begin{document}

\title[A Distribution Law for CCS and a New Congruence Result for
  the $\pi$-calculus]{A Distribution Law for CCS\\ 
  and a New Congruence Result for the $\pi$-calculus\rsuper *}

\author{Daniel Hirschkoff \and Damien Pous}
\email{\{Daniel.Hirschkoff,Damien.Pous\}@ens-lyon.fr}
\address{ENS Lyon, Universit{\'e} de Lyon, LIP (UMR 5668 CNRS ENS Lyon
  UCBL INRIA), France}

\keywords{Bisimulation, process algebra, CCS, $\pi$-calculus,
  axiomatisation} 
\subjclass{F.e.2}
\titlecomment{{\lsuper *} A preliminary version of this work appeared
  in~\cite{HP:fossacs07}} 

\begin{abstract}
  We give an axiomatisation of strong bisimilarity on a small fragment
  of CCS that does not feature the sum operator. This axiomatisation
  is then used to derive congruence of strong bisimilarity in the
  finite $\pi$-calculus in absence of sum.
  To our knowledge, this is the only nontrivial subcalculus of the
  $\pi$-calculus that includes the full output prefix and for which
  strong bisimilarity is a congruence.
\end{abstract}

\maketitle



\section*{Introduction}
\input{intro}

\section{MicroCCS Processes and Normal Forms}
\label{ssec:ccs}
\input{ccs}

\section{Characterisation of Bisimilarity in MicroCCS}
\label{ssec:bisim:ccs}
\input{otherproof}

\section{Nonexistence of a Finite Axiomatisation}
\label{ssec:nofinite}

\input{nofinite_long}

\section{On Substitution Closure of Bisimilarity}
\label{sec:distributed}
\input{dist}

\section{A New Congruence Result for the \texorpdfstring{$\pi$}{pi}-calculus}
\label{sec:pi}
\input{pi}

\section{Conclusion}
\label{sec:concl}
\input{concl}

\bigskip

\paragraph{Acknowledgements.} We are grateful to Arnaud Carayol
for interesting discussions at early stages of this work, and to
Ilaria Castellani for discussions about distributed bisimilarity.
An anonymous FOSSACS referee provided numerous suggestions, which
helped us in particular to improve the proof of
Theorem~\ref{thm:carac:norm:ccs}.
We benefited from support by the french initiative ``ACI GEOCAL'' and
from the ANR projects ``MoDyFiable'' and ``CHoCO''.

\bibliography{refs.bib}
\bibliographystyle{plain}

\end{document}

%% file: intro.tex
{ In this paper, we study strong bisimilarity on two process
  calculi.  More precisely, we establish an axiomatisation for strong
  bisimilarity on a very restricted fragment of CCS, and then use this
  axiomatisation to derive a new congruence result for the
  $\pi$-calculus.  }

\medskip

We first focus on \emph{microCCS} (\microccs), the subcalculus of CCS
that only features prefix and parallel composition.  Our main result
on \microccs{} is that adding the following \emph{distribution law}
\begin{maths}
  \eta.(P\|\eta.P\|\dots\|\eta.P)\quad =\quad
  \eta.P\|\eta.P\|\dots\|\eta.P
\end{maths}
\noindent to the laws of an abelian monoid for parallel composition
yields a complete axiomatisation of strong bisimilarity (in the law
above, $\eta$ is a CCS prefix, of the form $a$ or $\overline{a}$, and
$P$ is any CCS process -- the same number of copies of $P$ appear on
both sides of the equation). 

The distribution law is not new: it is mentioned -- among other
\emph{`mixed equations'} relating prefixed terms and parallel
compositions -- in a study of bisimilarity on normed PA
processes~\cite{hirshfeld:jerum:tech}.
In our setting, this equality can be oriented from left to right to
rewrite processes into normal forms, which intuitively exhibit as much
concurrency as possible. Strong bisimilarity ($\sim$) between
processes is then equivalent to equality of their normal forms.
This rewriting phase allows us to actually compute \emph{unique
  decompositions} of processes into \emph{prime processes}, in the
sense of~\cite{milnermoller}: a process $P$ is prime if $P$ is not
bisimilar to the inactive process $\nil$ and if $P\sim Q\|R$ implies
$Q\sim\nil$ or $R\sim\nil$.

The distribution law is an equational schema, corresponding to an
infinite family of axioms, of the form $\eta.(P\|(\eta.P)^{k}) =
(\eta.P)^{k+1}$, for $k\geq 1$ (where $Q^k$ denotes the $k$-fold
parallel composition of process $Q$). 
{ Some of these axioms are related. For instance, we can derive
  the 6-ary instance of the distribution law (corresponding to $k=5$)
  using the binary and the ternary instances: first rewrite
  $(\eta.P)^6$ three times using the binary instance, yielding
  $(\eta.(P|\eta.P))^3$; then use the ternary instance to rewrite the
  latter process into $\eta.\big(\,(P|\eta.P)~|~\eta.(P|\eta.P)~|~
  \eta.(P|\eta.P)\big)$; finally, use the binary instance twice to get
  $\eta.(P\,|\,(\eta.P)^5)$.
  On the other hand, instances of the distribution law where the
  prefixed term occurs a prime number of times on the right hand side
  cannot be derived using other instances. We formalise this argument
  to show that there exists no finite axiomatisation of $\sim$ on
  \microccs{} in Sect.~\ref{ssec:nofinite}.  }

\medskip

We are also interested in this paper in the $\pi$-calculus, and, more
precisely, in congruence properties of strong bisimilarity in this
formalism. Because of the presence of the input prefix, and of the
related phenomenon of name-passing, bisimilarity is more complex in
the $\pi$-calculus than in CCS\@. In particular, both early and late
bisimilarity~\cite{SW01}, that differ in their treatment of name
substitution, fail to be congruences in the full $\pi$-calculus.

There exist subcalculi of the $\pi$-calculus for which strong
bisimilarity is a congruence ({we discuss these in
  Sect.~\ref{sec:concl}}). When this is the case, this equivalence
coincides with \emph{ground bisimilarity} (\simg), which allows one to
consider a single fresh name when analysing an input transition,
instead of the usual quantification involving all free names of the
process. Congruence of strong bisimilarity is hence an important
property: not only is it necessary in order to reason in a
compositional way, but it also helps making bisimulation proofs
simpler, by reducing the number of cases to analyse.

In the full $\pi$-calculus, in order to get congruence, one has to
work with Sangiorgi's open bisimilarity~\cite{sangiorgi:open}, which
has a more involved definition than the early and late variants. Tools
like the Mobility Workbench~\cite{MWB}, for instance, have adopted
this equivalence on processes. 

\medskip

Technically, the key property which is necessary in order to derive
congruence of \simg{} in the $\pi$-calculus is \emph{substitution
  closure}: we say that a relation \RR{} between processes is closed
under substitution if whenever $P\RR Q$, then $P\sigma\RR Q\sigma$ for
any substitution $\sigma$ mapping names to names. 
In calculi like CCS or the $\pi$-calculus, where interaction arises
from the synchronisation between an emitter and a receiver,
substitution closure is a demanding property. Indeed, applying a
substitution may have the effect of identifying two names, thus
triggering new possibilities of interaction.

Before addressing substitution closure for \simg{} in the
$\pi$-calculus, we analyse this property in the simpler setting of
(subsets of) CCS in Sect.~\ref{sec:distributed}. We show in particular
that strong bisimilarity is closed under substitution in \microccs,
but that it is \emph{not} as soon as we add the choice operator,
although being a congruence. 


At the heart of our proof of congruence in the $\pi$-calculus is a
notion that we call \emph{\MD}, and that corresponds to the existence
of processes $P, P_{12}, P_{21}$ such that
$P\xr{\eta_1}\xr{\eta_2}P_{12}$ and $P\xr{\eta_2}\xr{\eta_1}P_{21}$,
for two distinct actions $\eta_1$ and $\eta_2$, and with $P_{12}$
behaviourally equivalent to $P_{21}$.
(We do not specify the shape of actions, nor the behavioural
equivalence we refer to, because we shall be reasoning about \MD s both
in \microccs{} and in the $\pi$-calculus.)
We additionally require in the two sequences of transitions from $P$
to $P_{12}$ and $P_{21}$ respectively that the second prefix being
fired should occur under the first prefix in $P$.
%
%

We discuss the relationship between substitution closure and \MD{}s in
Sect.~\ref{sec:MDs}, and show that the latter do not arise in
\microccs{} (which is a way to prove that $\sim$ is closed under
substitution in this calculus).  This is essentially due to the fact
that our axiomatisation of $\sim$ on \microccs{} does not allow one to
match the firing of two \emph{distinct} prefixes that are concurrent
using two prefixes that occur in sequence in a process.

In relation with the latter observation, we then argue in
Sect.~\ref{sec:noninterleaving} that \emph{noninterleaving semantics},
for which concurrency cannot be reduced to nondeterminism, are more
likely to be substitution closed: we prove that this is the case for
Castellani and Hennessy's \emph{distributed
  bisimilarity}~\cite{DBLP:journals/jacm/CastellaniH89} in \microccs{}
extended with choice.

\medskip

Coming back to the $\pi$-calculus, we exploit a transfer property
that allows us to derive from the absence of \MD{}s in \microccs{} the
same result in \PI, the finite, sum-free $\pi$-calculus. This entails
that ground, early, late and open bisimilarities coincide on \PI, and
are congruences.
It is known~\cite{SW01} that bisimilarity in the $\pi$-calculus fails
to be a congruence as soon as we have prefix, parallel composition,
restriction and replication. 
The problem of congruence of \simg{} on \PI{} is mentioned as an open
question in~\cite[Chapter 5]{SW01}, and is known since at least
1998~\cite{boreale:sangiorgi:congruence}. To our knowledge, this is
the first congruence result for a subcalculus of the $\pi$-calculus
that includes the full output prefix (see Sect.~\ref{sec:concl} for a
discussion on this).

%


\iflong

\rouge{
To establish congruence of ground bisimilarity (\simg), it is necessary to
show that this equivalence is closed under substitutions, i.e., that
whenever $P\simg Q$, then $P\sigma\simg Q\sigma$ for any substitution
$\sigma$ mapping names to names. We call this property
\emph{substitution closure} of \simg.
\\
Let us look at the proof that \simg{} is closed under substitutions.
For this, we reason by coinduction, and consider a transition
$P\sigma\xr{\mu}P'$. If $\mu$ is a visible action (an emission or a
reception), we rather easily show that this transition corresponds to
a transition along some action $\mu_0$ of $P$, which by hypothesis is
matched by $Q$ (i.e., $Q\xr{\mu_0}Q'_0$ for some $Q'_0$), and we can
finally deduce $Q\sigma\xr{\mu}Q'$ with $P'\simg Q'$. A similar
reasoning can be made if $P\sigma$ moves along a $\tau$ transition
that corresponds to an original $\tau$ transition of $P$.
\\
The interesting case is when $P$ is liable to perform an emission and
a reception on two different names, that have the same image by
$\sigma$. That is, $P\xr{a(x)}P'_1$, $P\xr{\out{b}{p}}P'_2$, and
$\sigma(a)=\sigma(b)$, from which we deduce $P\sigma\xr{\tau}P'$ --
we work here with the \emph{late} transition system for the
$\pi$-calculus (see Sect.~\ref{sec:pi}). 
Because $P\simg Q$, we know $Q\xr{a(x)}Q'_1$ and
$Q\xr{\out{b}{p}}Q'_2$, with $P'_i\simg Q'_i,\, i=1,2$. Since we work
in a sum-free calculus, we can define $Q'$ such that
$Q\sigma\xr{\tau}Q'$, but do we have $P'\simg Q'$?
\\
It is known~\cite{SW01} that bisimilarity in the $\pi$-calculus fails
to be a congruence as soon as we have prefix, parallel composition,
restriction and replication. The counterexample is the following: take
\begin{mathpar}
  P_0~\eqdef~ !a.\out{b}.\tau.p~|~ !\out{b}.a.\tau.p
  \and
  \mbox{and}
  \and
  Q_0~\eqdef~ !(\new c)\,(a.\out{c}~|~\out{b}.c.p)
\end{mathpar}
\noindent (this example is actually phrased in CCS with restriction
  and replication; the construction $\tau.P$ can be encoded as $(\new
  d)\,(d.P|\out{d})$, for some fresh channel name $d$).
  We have $P_0\simg Q_0$, but if we define $\sigma$ as the
  substitution that replaces $b$ with $a$, we have $P_0\sigma\not\simg
  Q_0\sigma$, because one process is liable to do two synchronisations
  and interact on $p$, while the other one needs at least three
  synchronisations to do so.
}
\fi

\paragraph{Paper outline.}

We introduce \microccs{} and the distribution law in
Sect.~\ref{ssec:ccs}. Section~\ref{ssec:bisim:ccs} is devoted to the
characterisation of $\sim$ on \microccs{} using normal forms.  In
Sect.~\ref{ssec:nofinite}, we prove that no finite axiomatisation of
$\sim$ on \microccs{} exists.  
We discuss the substitution closure property, and establish it for
distributed bisimilarity in an extension of \microccs, in
Sect.~\ref{sec:distributed}.
Section~\ref{sec:pi} presents the proof
of our congruence result in the $\pi$-calculus, and we give concluding
remarks in Sect.~\ref{sec:concl}.

This paper is an extended version of~\cite{HP:fossacs07}. In
particular, we provide more detailed proofs in
Sect.~\ref{ssec:nofinite}; the material in
Sect.~\ref{sec:noninterleaving}, that discusses substitution closure
and noninterleaving semantics, is new.


%% file: ccs.tex
We consider an infinite set \names{} of names, and let $a,b\dots$
range over names. We define on top of \names{} the set of processes of
\microccs, the finite, public (that is, without restriction), sum-free
CCS calculus, as follows, where $P,Q,R\dots$ range over processes:
\begin{align*}
  \eta &::= a \OR \overline{a}\enspace, & %
  P &::= \nil \OR \eta.P \OR P\|Q\enspace.
\end{align*}
\nil{} is the nil process. $\eta$ ranges over \emph{interactions}
(also called \emph{visible actions}), and we let $\overline{\eta}$
stand for the \emph{coaction} associated to $\eta$ (we let
$\overline{\overline{\eta}} = \eta$). For $k>0$, we write $P^k$ for
the parallel composition of $k$ copies of $P$, and we write
$\prod_{i\in I}P_i$ for the parallel composition of all processes
$P_i$ for $i\in I$.

\emph{Structural congruence}, written $\equiv$, is defined as the
smallest congruence satisfying the following laws:
\begin{align*}
  (C_1)&\quad P\|Q \equiv Q\|P &\quad%
  (C_2)&\quad P\|(Q\|R) \equiv (P\|Q)\|R &\quad%
  (C_3)&\quad P\|\nil \equiv P
\end{align*}
We introduce a labelled transition system (LTS) for \microccs.
\emph{Actions} labelling transitions are either interactions, or a
special \emph{silent action}, written $\tau$. We use $\mu$ to range
over actions.
It can be noted that the syntax of \microccs{} does not include a
construction of the form $\tau.P$ -- see Remark~\ref{rk:weak} below.

\begin{defi}[Operational semantics and behavioural equivalence]
  \label{def:ccs:sos}~\\
  The LTS for \microccs{} is given by the following rules:
  \begin{maths}
    \infer{}{\eta.P \xr{\eta} P}
    \qquad
    \infer{P\xr{\eta}P'\qquad Q\xr{\overline{\eta}}Q'}{P\|Q\xr{\tau}P'\|Q'}
    \qquad
    \infer{P\xr{\mu}P'}{P\|Q\xr{\mu}P'\|Q}
    \qquad
    \infer{P\xr{\mu}P'}{Q\|P\xr{\mu}Q\|P'}
  \end{maths}
  A \emph{bisimulation} is a symmetrical relation $\RR$ between
  processes such that whenever $P\RR Q$ and $P\xr\mu P'$, there
  exists $Q'$ such that $Q\xr\mu Q'$ and $P'\RR Q'$.
  
  \noindent
  \emph{Bisimilarity}, written $\sim$, is the union of all
  bisimulations.
\end{defi}
\begin{defi}[Size]
  Given $P$, \size{P} (called the \emph{size} of $P$) is defined by:
  \begin{align*}
    \size{\nil}&\eqdef 0 & %
    \size{P_1\|P_2}&\eqdef \size{P_1}+\size{P_2} & %
    \size{\eta.P}&\eqdef 1+\size P\enspace.
  \end{align*}
\end{defi}
\begin{lem}\label{lem:bisim_size}
  $P\equiv Q$ implies $P\sim Q$ which in turn implies
  $\size{P}=\size{Q}$.
\end{lem}
\begin{proof}
  The first implication follows by showing that the laws of $\equiv$
  are sound for $\sim$, and that $\sim$ is preserved by parallel
  composition and prefix.
  
  Assume then by contradiction that there exist $P,Q$ such that
  $P\sim Q$ and $\size{P}<\size{Q}$; and choose such $P$ with minimal
  size.  $Q$ has at least one prefix: $Q\xr\eta Q'$ and we get
  $P\xr\eta P'$ with $P'\sim Q'$. We deduce that $\size{P'}<\size{P}$
  and $\size{P'}<\size{Q'}$, which contradicts the minimality
  hypothesis.  
\end{proof}
\begin{defi}[Distribution law]\label{def:distriblaw}
  The \emph{distribution law} is given by the following equation,
  where the same number of copies of $P$ appears on both sides:
  \begin{maths}
    \eta.(P\|\eta.P\|\dots\|\eta.P)\quad=\quad
    \eta.P\|\eta.P\|\dots\|\eta.P\enspace.
  \end{maths}
  We shall use this equality, oriented from left to right, to rewrite
  processes.  We write $P\rewd{}P'$ when there exist $P_1, P_2$ such
  that $P\equiv P_1$, $P_2\equiv P'$ and $P_2$ is obtained from $P_1$
  by replacing a sub-term of the form of the left-hand side process
  with the right-hand side process.
\end{defi}

\begin{remark}[On the distribution law and PA]
  Among the studies about properties of $\sim$ in process algebras
  that include parallel composition (see~\cite{aceto:fokkink:survey}
  for a recent survey on axiomatisations), some works focus on calculi
  where parallel composition is treated as a primitive operator (as
  opposed to being expressible using sum or other constructs like the
  left merge operator).  As mentioned above, particularly relevant to
  this work is~\cite{hirshfeld:jerum:tech}, where Hirshfeld and
  Jerrum \textit{``develop a structure theory for PA that completely
    classifies the situations in which a sequential composition of two
    processes can be bisimilar to a parallel composition''}.
  \cite{hirshfeld:jerum:tech} establishes decidability of $\sim$
  for normed PA processes: in that setting, the formal analogue of the
  distribution law (Def.~\ref{def:distriblaw}) holds with $\eta$ and
  $P$ being two processes -- the `dot' operator is a general form of
  sequential composition.  This equality is valid
  in~\cite{hirshfeld:jerum:tech} whenever $\eta$ is a `monomorphic
  process', meaning that $\eta$ can only reduce to \nil{} (which
  corresponds to \microccs), or to $\eta$ itself.
  \cite{fokking:luttik:icalp00} presents a finite axiomatisation of PA
  that exploits the operators of sum and left merge.
\end{remark}

\begin{lem}\label{lem:confl}  
  The relation \rewd{} is strongly normalising and confluent.
\end{lem}
\begin{proof}
  If $P\rewd P'$ then the weight of $P'$ (defined as sum of the depths
  of all prefixes occurring in $P'$) is strictly smaller than the
  weight of $P$, whence the strong normalisation.
  We then remark that $\rewd$ is locally confluent, and conclude with
  Newman's Lemma.
\end{proof}
Thus, for any process $P$, \rewd{} defines a normal form unique up to
$\equiv$, that will be denoted by $\norm P$. We let $A,B,\dots$ range
over normal forms.

The following lemma states that $\rewd$ preserves bisimilarity:
\begin{lem}\label{lem:rwd:bisim}
  If $P\rewd P'$, then $P\sim P'$. For any $P$, $P\sim\norm{P}$.
\end{lem}
\begin{proof}
  The relation $(\s\rewd\cup\s{(\rewd)^{-1}}
  \cup\s\equiv)$ is a bisimulation.
\end{proof}



%% file: otherproof.tex
Our characterisation of $\sim$ on \microccs{} makes use of the notion
of decomposition into \emph{prime processes}, defined as follows:

\begin{defi}\label{def:prime}
  A process $P$ is \emph{prime} if $P\not\sim\nil$ and $P\sim P_1\|P_2$
  implies $P_1\sim\nil$ or $P_2\sim\nil$.

  When $P\sim P_1\|\dots\|P_n$ where the $P_i$s are prime, we shall
  call $P_1\|\dots\|P_n$ a \emph{prime decomposition} of $P$.
\end{defi}

\begin{prop}[Unique decomposition]\label{prop:uniquedecomp}
  Any process admits a prime decomposition which is unique up to
  bisimilarity: 
  if $P_1\|\dots\|P_n$ and $Q_1\|\dots\|Q_m$ are two prime
  decompositions of the same process, then $n=m$ and there exists a
  permutation $f$ of $[1..n]$ such that $P_i\sim Q_{f(i)}$ for all
  $i\in[1..n]$.
\end{prop}
\begin{proof}
  Similar to the proof of~\cite[Theorem~4.3.1]{moller:phd}: the case
  of \microccs{} is not explicitly treated in that work, but the proof
  can be adapted rather easily. 
\end{proof}

An immediate consequence of the above result is the following
property:

\begin{cor}[Cancellation]\label{coro:cancellation}
  For all $P,Q,R$, $P\|R\sim Q\|R$ implies $P\sim Q$.
\end{cor}
Note that this is not true in presence of replication:
$a\|!a\sim\nil\|!a$, but $a\not\sim\nil$.

\medskip

The characterisation of $\sim$ using the distribution law follows from
the observation that if a normal form is a prefixed process, then it
is prime. This idea is used in the proof of Lemma~\ref{lem:simequiv}.
We first establish a technical result, that essentially exploits the
same argument as the proof of Theorem~4.2
in~\cite{hirshfeld:jerum:tech}.

\begin{lem}\label{lem:mixedeq}
  If $\eta.P\sim Q\|Q'$, with $Q,Q'\not\sim\nil$, then there exist $A$
  and $k>1$ such that $\eta.P\sim (\eta.A)^k$ and $\eta.A$ is a normal
  form.
\end{lem}
\begin{proof}
  By Lemma~\ref{lem:rwd:bisim}, we have $\eta.P\sim \norm{Q\|Q'}$.
  Furthermore, we have that $\norm{Q\|Q'} \equiv
  \prod_{i\le k}\eta_i.A_i$, where $k>1$ and the processes $\eta_i.A_i$
  are in normal form.

  Since the $\eta$ prefix must be triggered to answer any challenge
  from the right hand side, we have $\eta_i=\eta$ and $P\sim
  A_i\|\prod_{l\neq i}\eta.A_l$ for all $i\le k$.
  In particular, when $i\neq j$, we have $P\sim
  A_i\|\eta.A_j\|\prod_{l\not\in \{i,j\}}\eta.A_l\sim
  \eta.A_i\|A_j\|\prod_{l\not\in \{i,j\}}\eta.A_l$ and hence, by
  Corollary~\ref{coro:cancellation}, $A_i\|\eta.A_j\sim \eta.A_i\|A_j$.
  By reasoning on the sizes of the parallel components in the prime
  decompositions of these two terms, we conclude that $\eta.A_i\sim
  \eta.A_j$ for all $i,j\le k$.
  
  Hence, we have $\eta.P\sim(\eta.A_1)^k$ with $k>1$ and $\eta.A_1$ is
  a normal form. 
\end{proof}

\begin{lem}\label{lem:simequiv}
  Let $A,B$ be two normal forms, $A\sim B$ implies $A\equiv B$.
\end{lem}
\proof
  We show by induction on $n$ that for all $A$ with $\size{A}=n$, we
  have
  \begin{enumerate}[(i)]
  \item if $A$ is a prefixed process, then $A$ is prime;
  \item for any $B$, $A\sim B$ implies $A\equiv B$.
  \end{enumerate}
  The case $n=0$ is immediate. Assume that the property holds for all
  $i<n$, with $n\geq 1$.
  \begin{enumerate}[(i)]
  \item We write $A=\eta.A'$, and assume by contradiction
    $A\sim P_1\|P_2$ with $P_1,P_2\not\sim \nil$. By
    Lemma~\ref{lem:mixedeq}, we have $A\sim (\eta.B)^k$ with $k>1$ and
    $\eta.B$ in normal form.  By triggering the prefix on the left
    hand side, we have $A'\sim B\|(\eta.B)^{k-1}$. It follows by
    induction that $A'\equiv B\|(\eta.B)^{k-1}$ (using property
    ($ii$)), and hence $A\equiv \eta.(B\|(\eta.B)^{k-1}$, which is in
    contradiction with the fact that $A$ is in normal form.

  \item Assume now $A\sim B$. 
    \begin{enumerate}[$-$]
    \item If $A$ is a prefixed process, $B$ is prime by the previous
      point ($\size B=\size A$ by Lemma~\ref{lem:bisim_size}).
      Necessarily, $ A\equiv \eta.A'$ and $B\equiv \eta.B'$ with
      $A'\sim B'$. By induction, this entails $A'\equiv B'$, and
      $A\equiv B$.
    \item Otherwise, $A=\eta_1.A_1\|\dots\|\eta_k.A_k$ with $k>1$, and
      we know by induction (property ($i$)) that $\eta_i.A_i$ is prime
      for all $i\le k$.  Similarly, we have
      $B=\eta'_1.B_1\|\dots\|\eta'_l.B_l$ with $\eta'_i.B_i$ prime for
      all $i\le l$.
      
      By Proposition~\ref{prop:uniquedecomp}, $k=m$ and
      $\eta_i.A_i\sim \eta'_i.B_i$ (up to a permutation of the
      indices), which gives $\eta'_i=\eta_i$ and $A_i\sim B_i$ for all
      $i\leq k$.
      By induction, we deduce $A_i\equiv B_i$ for all $i$, which
      finally implies $A\equiv B$.\qed
    \end{enumerate}
  \end{enumerate}

\noindent Lemmas~\ref{lem:rwd:bisim} and~\ref{lem:simequiv} allow us to deduce
the following result.

\begin{thm}\label{thm:carac:norm:ccs}
  Let $P, Q$ be two \microccs{} processes. Then $P\sim Q$ iff
  $\norm{P}\equiv \norm{Q}$.
\end{thm}

\begin{remark}[Unique decomposition of processes]
  Our proof relies on unique decomposition of processes
  (Prop.~\ref{prop:uniquedecomp}), that first appeared
  in~\cite{milnermoller}. Unique decomposition has been established
  for a variety of process algebras, and used as a way to prove
  decidability of behavioural equivalence and to give complexity
  bounds for the associated decision procedure
  (\cite{luttik:concurrency:column,burkart:caucal:moller:steffen:handbook}
  cite relevant references).
  
  In the present study, beyond the existence of a unique
  decomposition, we are interested in a syntactic characterisation of
  $\sim$ (which will in particular allow us to derive
  Lemma~\ref{lem:grange} below).
  In this sense, our work is close
  to~\cite{corradini:gorrieri:marchignoli:1998}, where the notion of
  \emph{maximally parallel process} in CCS (with choice) is studied.
  \cite{corradini:gorrieri:marchignoli:1998} defines a rewriting
  process through which maximally parallel normal forms can be
  computed, and shows that in the case of \microccs, such normal forms
  are unique.  However, no syntactical characterisation of the set of
  normal forms is presented, and such a characterisation cannot be
  directly deduced from the (rather involved) definition of the
  rewriting process for full CCS.
  
  We instead restrict ourselves to \microccs{} from the start, and
  rely explicitly on the distribution law in order to `extract' prime
  components of processes.
\end{remark}

  

\begin{remark}[$\tau$ prefix and weak bisimilarity]
  \label{rk:weak}
  We do not address weak bisimilarity in the present work.  In
  \microccs, strong and weak bisimilarity coincide, i.e., the internal
  transitions of processes are completely determined by the visible
  actions (interactions).  This is essentially due to the absence of
  restriction in the calculus.
  When including $\tau$ prefixes in the syntax, it can be proved that
  adding the law $\tau.P=P$ is enough to characterise weak
  bisimilarity.
  The $\tau$ prefix is usually absent in the $\pi$-calculus, to which
  we shall move in Sect.~\ref{sec:pi}.
  Since  some results on CCS will be transferred to the $\pi$-calculus,
  we did not include this construct in \microccs.
\end{remark}


%% file: nofinite_long.tex
We let $M,N$ range over \microccs{} terms with \emph{variables} (this
corresponds to the grammar $M ::= \nil\OR\eta.M\OR M|M\OR X$, and we
use $X,Y\dots$ to range over term variables).
A \emph{ground term} is a term with no occurrence of variables.
\emph{Instantiations} are mappings from variables to terms, and their
domain are naturally extended to terms.  We use \gsub{} to range over
instantiations.  Applying \gsub{} to $M$ yields a term written
$M\gsub$.  \gsub{} is a \emph{ground instantiation} if for all terms
$M$, $M\gsub$ is a ground term.
  Any two terms $M,N$ define an \emph{equation}, written $M=N$.

\begin{defi}[Axiomatic equality]
  Given a set \EE{} of equations, we shall write \deriv\EE M N
  whenever $M=N$ can be derived in equational logic using equations
  from \EE.
  
  We let $\DD$ stand for the set of equations consisting of the three
  axioms of structural congruence $(C_1, C_2, C_3)$, and all the
  distribution axioms $((D_i)_{i\ge 1})$:
  \begin{maths}
    (D_i):~\eta.(P\|(\eta.P)^i)~=~(\eta.P)^{i+1},~i\geq 1\enspace.
  \end{maths}
  $\DD_k$ stands for the
  finite restriction of $\DD$ where only the first $k$ distribution
  axioms are included $((D_i)_{1\le i\le k})$.
\end{defi}

Equations of \DD{} are obviously sound for $\sim$. Ground completeness
is given by the following proposition, which holds by
Theorem~\ref{thm:carac:norm:ccs}.
\begin{prop}[Completeness]\label{prop:complete}
  For any processes $P,Q$, 
  \begin{maths}
    P\sim Q\quad\textrm{iff}\quad\deriv \DD P Q\enspace.
  \end{maths}
\end{prop}

We now analyse the distribution law using a rather classical
approach~\cite{aceto:fokkink:survey}.
We show that \DD{} is $\omega$-complete, that is, complete w.r.t.\ the
extensional equality derived from strong bisimilarity.  Since, by
Lemma~\ref{lem:primes} below, \DD{} is \emph{intrinsically} infinite,
we derive impossibility of a finite axiomatisation of $\sim$ on
\microccs, by using compactness arguments.
\begin{defi}[Extensional equality]
  Two terms $M$ and $N$ are \emph{extensionally equal}, written \ext M
  N, whenever for any ground instantiation \gsub, it holds that
  $M\gsub\sim N\gsub$. An equation $M=N$ is said to be \emph{correct}
  if \ext M N.
\end{defi}

Our proof of $\omega$-completeness essentially relies on the
methodology developped in~\cite{Groote90}; the idea is to replace
variables by small terms that can easily be distinguished.
\begin{lem}\label{lem:normsubst}
  Let $M$ be a term whose variables all belong to $\{X_i\}_{i\in I}$,
  and let $\{a_i\}_{i\in I}$ be a collection of distinct names that do
  not occur in $M$.
  \begin{maths}
    \norm{M\{a_i.0/X_i\}}\equiv \norm{M}\{a_i.0/X_i\}
  \end{maths}
\end{lem}
\begin{proof}
  We proceed by well founded induction over the termination of \rewd{}.
  \begin{enumerate}[$\bullet$]
  \item If $M$ is in normal form, we just have to check that
    $M\{a_i.0/X_i\}$ is in normal form. This is true because the $a_i$
    are distinct and do not appear in $M$.
  \item Otherwise, if $M\rewd N$, we check that $M\{a_i.0/X_i\}\rewd
    N\{a_i.0/X_i\}$ so that:
    \begin{maths}
      \begin{array}{rcl@{\qquad}r}
        \norm{M\{a_i.0/X_i\}}&\equiv& \norm{N\{a_i.0/X_i\}}& \textrm{
          (by confluence)}\\
        &\equiv& \norm{N}\{a_i.0/X_i\}&\textrm{ (by induction)} \\
        &\equiv& \norm{M}\{a_i.0/X_i\}&\textrm{ (by confluence)} \\
      \end{array}
    \end{maths}
  \end{enumerate}
\end{proof}

\begin{lem}\label{lem:psub} Let $M, N$ be two terms whose variables 
  all belong to $\{X_i\}_{i\in I}$, and let $\{a_i\}_{i\in I}$ be a
  collection of distinct names that do not occur in $M$ nor in $N$.
  \begin{enumerate}[$\bullet$]
  \item If $\deriv \DD M N$ then $\deriv \DD {M\gsub} {N\gsub}$ for
    any instantiation \gsub;
  \item if $M\{a_i.0/X_i\}\sim N\{a_i.0/X_i\}$ then $\deriv \DD M N$.
  \end{enumerate}
\end{lem}
\begin{proof}
  The first point is standard, and proved by induction over the
  derivation tree.
   
  For the second property, we know by Theorem~\ref{thm:carac:norm:ccs}
  that $\norm{M\{a_i.0/X_i\}}\equiv \norm{N\{a_i.0/X_i\}}$. By
  Lemma~\ref{lem:normsubst}, we can deduce
  $\norm{M\{a_i.0/X_i\}}\equiv \norm{M}\{a_i.0/X_i\}$, and
  $\norm{N\{a_i.0/X_i\}}\equiv \norm{N}\{a_i.0/X_i\}$. Hence we have
  $\norm M\equiv \norm N$, and
  $\deriv \DD M N$ holds.
\end{proof}

\begin{thm}[$\omega$-completeness]\label{thm:omega}
  For any terms $M, N$, 
  \begin{maths}
    \ext{M}{N}\quad\textrm{iff}\quad\deriv{\DD}{M}{N}\enspace.
  \end{maths}
\end{thm}
\begin{proof}
  Using Lemma~\ref{lem:psub}, $\omega$-completeness boils down to the
  completeness of \DD{} for ground terms (Prop.~\ref{prop:complete}).
\end{proof}

Notice that the proof of Theorem~\ref{thm:omega} relies on the
existence of an infinite number of names.
The following result is standard.

\begin{lem}[Compactness]
  \label{lem:compact}
  For any terms $M, N$, 
  \begin{maths}
    \deriv\DD M N\quad\textrm{iff}\quad\deriv{\DD_k} M N\textrm{ for
      some }k\enspace.
  \end{maths}
\end{lem}
\begin{proof}
  Equational proofs are finite objects.
\end{proof}

\begin{lem}\label{lem:primes}
  Let $a$ be a name, for any number $k$, there exists $n$ such that:
  \begin{maths}
    \nderiv{\DD_k}{a.a^n}{a^{n+1}}\enspace.
  \end{maths}
\end{lem}
Remember that $a^n$ stands for the $n$-ary parallel composition of
$a.\nil$, so that this equality is an instance of axiom $(D_n)$.
\begin{proof}
  Let $n$ be a number strictly greater than $k$ such that $n+1$ is
  prime, and let $\theta(P,Q)$ denote the predicate: ``$P\sim Q\sim
  a^{n+1}$, $P\equiv a.P'$, and $Q\equiv Q_1|Q_2$ with
  $Q_1,Q_2\not\equiv \nil$''.
  
  Assume \deriv{\DD_k}{a.a^n}{a^{n+1}}, and consider the shortest
  proof of \deriv{\DD_k} P Q for some processes $P, Q$ such that
  either $\theta(P,Q)$ or $\theta(Q,P)$. Since
  $\theta(a.a^{n},a^{n+1})$ holds, such a minimal proof does exist.
  We reason about the last rule used in the derivation of this proof
  in equational logic.  For syntactic reasons, this cannot be
  reflexivity, a contextual rule, nor one of the structural congruence
  axioms. It can be neither symmetry nor transitivity, since otherwise
  this would give a shorter proof satisfying $\theta$.
  The only possibility is thus the use of one of the distribution
  axioms, say $D_i$ with $1\le i\le k$ and $a^{n+1}\sim Q\equiv
  (a.Q')^{i+1}$. By Lemma~\ref{lem:bisim_size}, since
  $\size{a^{n+1}}=n+1$, $i+1$ has to divide $n+1$. This is
  contradictory, because we have $2\le i+1\le k+1<n+1$, and $n+1$ is
  prime.  
\end{proof}

We can finally prove the nonexistence of a finite axiomatisation of
$\sim$ on \microccs. 
The proof corresponds to a standard application of the
\emph{Compactness Theorem}~\cite{aceto:fokkink:survey}.

\begin{thm}[No finite axiomatisation of $\sim$]
  \label{lem:nofinite}
  For any finite set of correct equations \EE, there exist processes
  $P$ and $Q$ such that $P\sim Q$ but \nderiv \EE P Q.
\end{thm}
\begin{proof}
  By correctness, for any equation $M=N$ in \EE, \ext M N. Hence, by
  $\omega$-completeness we can prove any equation of \EE{} using
  \DD{}. By Lemma~\ref{lem:compact}, and since \EE{} is finite, there
  exists $k$ such that $\DD_k\vdash\EE$. By Lemma~\ref{lem:primes},
  there exists $n$ such that $a.a^n \sim a^{n+1}$ and
  \nderiv{\DD_k}{a.a^n}{a^{n+1}}; and thus,
  \nderiv\EE{a.a^n}{a^{n+1}}.
\end{proof}



%% file: dist.tex
We now discuss the property of substitution closure of behavioural
equivalences in (subcalculi of) \ccs. In the $\pi$-calculus, because
of the input prefix, substitution closure is in general a necessary
condition for bisimilarity to be a congruence.
The notion of \MD{}, which we define in Sect.~\ref{sec:MDs}, allows us
to show that $\sim$ is closed under substitution in \microccs{}. This
notion will be used to establish substitution closure (and then
congruence) of \simg{} in \PI{} in Sect.~\ref{sec:pi}.
We analyse substitution closure in an extension of \microccs, both for
strong bisimilarity and distributed bisimilarity, in
Sect.~\ref{sec:noninterleaving} (the latter section is not technically
necessary to establish the result on \PI, and can therefore be
skipped).

\subsection{\majMD{}s}\label{sec:MDs}

In \microccs, $\sim$ is closed under substitution. One way to prove
that is to rely on the axiomatisation from Sect.~\ref{ssec:bisim:ccs}:
two processes related by an instance of the distribution law remain
equivalent when a substitution mapping names to names %
%
%
is applied (we can show in particular that for any substitution
$\sigma$, $\norm{P\sigma} \equiv \norm{\norm{P}\sigma}$).

Here, we derive this result using an alternative general pattern, that
corresponds to the proof of substitution closure of \simg{} in
Sect.~\ref{sec:pi}.  To understand how the notion of \MD{} arises, we
sketch the proof of substitution closure of $\sim$. Suppose for that
$P\sim Q$, and consider a substitution $\sigma$. To prove $P\sigma\sim
Q\sigma$, we reason by coinduction, and consider a transition
$P\sigma\xr{\mu}P_0$.  The difficult case arises when $\mu=\tau$, and
the synchronisation follows from $P\xr{a}P_1$, $P\xr{\out{b}}P_2$,
with $\sigma(a)=\sigma(b)$.
We observe that because we work in \microccs, the transitions of $P$
to $P_1$ and $P_2$ are necessarily offered by distinct parallel
components of $P$.  $P$ can therefore do a transition along $a$
followed by a transition along $\out{b}$ to some $P'$, to which $Q$
can answer since $P\sim Q$.  If $Q$ answers by firing two prefixes
that belong to different parallel components (`concurrent prefixes'),
we are done: we can infer a $\tau$ transition for $Q\sigma$, and
conclude using coinduction. If this is not the case (i.e., if the
$\out b$ prefix fired by $Q$ was guarded by the $a$ prefix), we
consider the sequence where $P$ performs the two transitions in the
reversed order, first $\out{b}$ then $a$, and reason similarly.
Therefore, the only case where we cannot conclude occurs when $Q$
matches both sequences of transitions using causally dependent
prefixes. This situation is depicted below; we will show that it
cannot arise in \microccs.

\begin{align*}
  \xymatrix { %
    &P\ar[rd]^{\out b}\ar[dl]_a \ar@{.}@/^2ex/[rrrr] &&&&Q
    \ar@{-->}[ldd] \ar@{-->}[rdd] \ar@{}[dd]|{a|\out b ?}
    \ar[rd]^{\out b}\ar[dl]_a\\
    P_1\ar[dr]_{\out b}&&P_2\ar[dl]^a
    &&\ar[d]_{\out b}&&\ar[d]^a\\
    &
    \ar@{.}@/_1ex/[rrr] \ar@{.}@/_3ex/[rrrrr]
    P'&&&Q_1&&Q_2\\
  }
\end{align*}
\medskip

More precisely, we show that the situation on the right of this
picture, where we notice that $Q_1\sim Q_2$ (both processes are
bisimilar to $P'$) cannot arise; we call such a -- hypothetical --
situation a \MD{}:


\begin{defi}[\majmd{} in \microccs]\label{def:MD}
  We say that there exists a \emph{\MD{} in \microccs} whenever
  there are two prefixes $\eta_1,\eta_2$, and five \microccs{}
  processes $S,S',T$, $T',R$ such that $\eta_1\neq\eta_2$,
  $S\xr{\eta_1}S'$, $T\xr{\eta_2}T'$ and
  $\eta_2.S\|T'\|R\sim S'\|\eta_1.T\|R$.
\end{defi}
%
%
We recover the situation which is depicted above by taking
$Q=\eta_2.S\|\eta_1.T\|R$, $\eta_1=a$, and $\eta_2=\out b$. Such a
notion is not specific to \microccs{}: the proofs of
Lemmas~\ref{lem:grange:pi:external} and~\ref{lem:grange:pi:internal}
will expose analogous situations in \PI.

\begin{defi}
  We define, for any \microccs{} process $P$ and prefix $\eta$, the
  \emph{contribution of $P$ at $\eta$}, written \sizeta{\eta}{P}, by
  \begin{align*}
    \sizeta{\eta}{\nil}&\eqdef 0 &
    \sizeta{\eta}{\eta'.P} &\eqdef 0\qquad\text{if }\eta\neq\eta' \\
    \sizeta{\eta}{P_1\|P_2} &\eqdef
    \sizeta{\eta}{P_1}+\sizeta{\eta}{P_2} &
    \sizeta{\eta}{\eta.P} &\eqdef \size{\eta.P}
  \end{align*}
\end{defi}
Intuitively, \sizeta{\eta}{P} is the total size of the parallel
components of $P$ that start with the prefix $\eta$.

\begin{lem}\label{lem:sizeta}
  $P\sim Q$ implies $\sizeta{\eta}{P} = \sizeta{\eta}{Q}$ for all
  $\eta$.
\end{lem}
\begin{proof}
  Follows from Theorem~\ref{thm:carac:norm:ccs} and the observation
  that the distribution law preserves the contribution of a process at
  a given interaction prefix. 
\end{proof}

\begin{lem}[No \MD]\label{lem:grange}
  There exists no \MD{} in \microccs.
\end{lem}
\begin{proof}
  Assume by contradiction that there are processes such that\
  $P\xr{\eta_1}P'$, $Q\xr{\eta_2}Q'$ and $\eta_2.P\|Q'\|R\sim
  P'\|\eta_1.Q\|R$.
  
  By the cancellation property (Corollary~\ref{coro:cancellation}), we
  have $\eta_2.P\|Q'\sim P'\|\eta_1.Q$, hence for all $\eta$,
  $\sizeta\eta{\eta_2.P\|Q'}=\sizeta\eta{P'\|\eta_1.Q}$
  (Lemma~\ref{lem:sizeta}).
  
  Since $\sizeta{\eta_1}{\eta_2.P\|Q'} = \sizeta{\eta_1}{Q'}\leq
  \size{Q'}$ and $\sizeta{\eta_1}{P'\|\eta_1.Q)} \geq
  \sizeta{\eta_1}{\eta_1.Q} = \size{Q'}+2$, by taking $\eta=\eta_1$ we
  finally get $\size{Q'}\geq\size{Q'}+2$.
\end{proof}

Lemma~\ref{lem:grange} will be used to show that a situation
corresponding to a \MD{} cannot arise in \PI{}.  Notice that the proof
depends in an essential way on Lemma~\ref{lem:sizeta}, which in turn
relies on the axiomatisation of $\sim$ in \microccs{}
(Theorem~\ref{thm:carac:norm:ccs}).

As a consequence of this result, we can deduce the following

\begin{cor}[Substitution closure of $\sim$ in \microccs]
  In \microccs, $P\sim Q$ entails $P\sigma\sim Q\sigma$, for all
  substitution $\sigma$.
\end{cor}

\medskip

We now introduce an extension of \microccs, called \microccsp, which
is the calculus obtained by adding a sum operator over prefixed
processes.  The grammar of \microccsp{} is thus the following:
\begin{mathpar}
  S~::=~ \nil\OR\eta.P\OR S_1+S_2
  \enspace,
  \and
  P~::=~ S\OR P_1|P_2
  \enspace.
\end{mathpar}
If $I=[1..k]$, we  write $\sum_{i\in I}S_i$ for $S_1+\dots+S_k$.
Like before, we use notation $\prod_i S_i$ for parallel compositions;
when using this notation, we shall moreover implicitly assume that for
all $i\in I$, $S_i{\not\dsim}\nil$ (this is in particular the case in
the statement of Lem.~\ref{lem:separation}).
We shall overload notations, and use $\sim$ to denote strong
bisimilarity in \microccsp.

\medskip

In \microccsp, $\sim$ is a congruence, but it is not closed under
substitution. We have indeed
\begin{equation}
  \label{eq:expansion}
  a\,|\,\out{b} ~\sim~ a.\out{b}+\out{b}.a
  \enspace.
\end{equation}
However, by applying the substitution that maps names $a$ and $b$ to
$p$, we obtain processes $p\,|\,\out{p}$ and $p.\out{p}+\out{p}.p$
respectively, which are not bisimilar: the former can do a $\tau$
transition that cannot be matched by the latter. 
Actually, $a.\out{b}+\out{b}.a$ gives a simple example of a \MD{} in
\microccsp.
This standard counterexample to substitution closure essentially
explains why early and late bisimilarities are not congruences in the
(full) $\pi$-calculus.

\begin{remark}[Restriction and replication instead of choice]
  As shown in~\cite{SW01}, a related counterexample can be constructed
  if, instead of adding the sum operator, we add restriction and
  replication to \microccs: the equivalence
\begin{mathpar}
  !a.\out{b}.\tau.q~|~ !\out{b}.a.\tau.q \quad\sim\quad !(\new
  c)\,(a.\out{c}~|~\out{b}.c.q)
\end{mathpar}
\noindent fails to hold if we replace $a$ and $b$ with $p$, because
one process is liable to do two synchronisations and interact on $q$,
while the other one needs at least three synchronisations to do so
(the construction $\tau.P$ can be encoded as $(\new
d)\,(d.P|\out{d})$, for some fresh channel name $d$).
\end{remark}

\subsection{Noninterleaving Semantics}\label{sec:noninterleaving}

We shall work in \microccsp{} in the remainder of this section.
It can be remarked that equality~(\ref{eq:expansion}) -- which is an
instance of the \emph{expansion law} -- is typical of
\emph{interleaving semantics}, in which the parallel composition of
two processes is equivalent to a single process, that expresses using
nondeterminism all possible interleavings of the two concurrent
activities.
As we have seen, equivalences that validate~(\ref{eq:expansion}), as
is the case for strong bisimilarity in \microccsp, are usually not
substitution closed.



On the contrary, we can expect locality-aware semantics, that are
sensitive to the parallel structure of processes (and hence more
discriminating than $\sim$), to be closed under substitution.
There are several approaches to define such equivalences.
We focus here on a version of (strong) \emph{distributed
  bisimilarity}~\cite{DBLP:journals/jacm/CastellaniH89,castellani:phd},
because it is among the simplest, and this will suffice for our
purposes.
The definition of distributed bisimilarity relies on \emph{distributed
  transitions}, which are given by judgements of the form
$P\dxr{\mu}\pair{P_1}{P_2}$. The intended meaning is that when $P$
performs the transition along $\mu$, it is decomposed into two parts.
At the site where the transition has happened, the local process
evolves into $P_1$ (the local residual). The remainder of the process,
which has not taken part in the transition, evolves into $P_2$ (the
concurrent residual). For example, we have
$P_1\,|\,\eta.Q\,|\,P_2\dxr{\eta}\pair{Q}{P_1|P_2}$\,.


The inference rules for distributed transitions in \microccsp{} are
the following (symmetrical versions of the rules for sum and parallel
composition are omitted):
\begin{mathpar}
  \inferrule*{}{\eta.P\dxr{\eta}\pair{P}{\nil}}
  \and
  \inferrule*{S\dxr{\eta}\pair{P_1}{P_2}}{S+S'\dxr{\eta}\pair{P_1}{P_2}}
  \and
  \inferrule*{P\dxr{\eta}\pair{P_1}{P_2}}{P|P'\dxr{\eta}\pair{P_1}{P_2|P'}}
  \and
  \inferrule*{P\dxr{\eta}\pair{P_1}{P_2}\and
    Q\dxr{\out{\eta}}\pair{Q_1}{Q_2}
    }{
    P|Q\dxr{\tau}\pair{P_1|Q_1}{P_2|Q_2}
    }
\end{mathpar}
%


\begin{defi}[Distributed bisimilarity]\label{def:dsim}
  A symmetric relation \RR{} between processes is a \emph{distributed
    bisimulation} iff whenever $P\RR Q$, if
  $P\dxr{\mu}\pair{P_1}{P_2}$, then there exist $Q_1, Q_2$ such that
  $Q\dxr{\mu}\pair{Q_1}{Q_2}$, $P_1\RR Q_1$ and $P_2\RR Q_2$.

Distributed bisimilarity, written \dsim, is the greatest distributed
bisimulation. 
\end{defi}


\begin{lem}\label{lem:dtrans}
  If $P\dxr{\eta}\pair{P_1}{P_2}$, then $P\equiv (\eta.P_1+S_1)~|~ P_2$
  for some $S_1$.
\end{lem}

\begin{lem}\label{lem:dsim:congr}
  \dsim{} is a congruence on \microccsp.
\end{lem}

\begin{prop}\label{prop:dsim:substclosed}
  \dsim{} is substitution closed in \microccsp.
\end{prop}

Prop.~\ref{prop:dsim:substclosed} is established by following the
reasoning we have sketched before Def.~\ref{def:MD}, but things are
considerably more easy due to distributed transitions, that insure
that concurrent prefixes can be fired.

\iflong
\begin{proof}
  We prove that $\RR~\eqdef~\set{(P\sigma,Q\sigma), P\dsim Q}$ is a
  distributed bisimulation. For that, we consider the transitions of
  $P\sigma$. The only difficult case is when $P\xr{\tau}P'$ because
  $P\xr{a}P'_1$, $P\xr{\out{b}}P'_2$, and $\sigma(a)=\sigma(b)$.
  These two transitions along $a$ and $\out{b}$ necessarily originate
  in distinct parallel components of $P$, since they lead to a
  synchronisation. This allows us to deduce, using
  Lemma~\ref{lem:dtrans}, $P\equiv (a.P_1+S_1)~|~(\out{b}.P_2+S_2)~|~P_3$,
  with $P\dxr{a}\pair{P_1}{(\out{b}.P_2+S_2)~|~P_3}$. The latter
  implies, since $P\dsim Q$, $Q\dxr{a}\pair{Q_1}{Q'_2}$ with $P_1\dsim
  Q_1$ and $(\out{b}.P_2+S_2)~|~P_3\dsim Q'_2$ for some $Q_1,Q'_2$.

  We have $(\out{b}.P_2+S_2)~|~P_3\dxr{\out{b}}\pair{P_2}{P_3}$, which
  implies $Q'_2\dxr{\out{b}}\pair{Q_2}{Q_3}$, for some $Q_2, Q_3$,
  with $P_2\dsim Q_2$ and $P_3\dsim Q_3$. But
  $Q'_2\dxr{\out{b}}\pair{Q_2}{Q_3}$ means $Q'_2\equiv
  (\out{b}.Q_2+S'_2)~|~Q_3$ for some $S'_2$ (Lemma~\ref{lem:dtrans}). 
  Therefore, we have $Q\equiv (a.Q_1+S'_1)~|~ (\out{b}.Q_2+S'_2)~|~ Q_3$,
  with $P_i\dsim Q_i$ for $i=1,2,3$. This gives by
  Lemma~\ref{lem:dsim:congr} $P_1|P_2|P_3\dsim Q_1|Q_2|Q_3$ and
  $Q\sigma\xr{\tau}Q_1\sigma\,|\, Q_2\sigma\,|\,Q_3\sigma$. By
  coinduction, we have $(P_1|P_2|P_3)\sigma\dsim(Q_1|Q_2|Q_3)\sigma$,
  which allows us to conclude.
%
\end{proof}
\fi

Actually, \dsim{} coincides with structural congruence in \microccsp{}
(in \microccsp, in addition to the equalities that are valid in
\microccs, $\equiv$ satisfies the laws of an abelian monoid for $+$,
as well as the idempotence law $S+S\equiv S$). To show this, we first
establish the following \emph{separation property}, enjoyed by \dsim{}
in \microccsp:

\begin{lem}[Separation Property]\label{lem:separation}
  If $P=\prod_{i\in I}S_i$, $Q=\prod_{j\in J}S'_j$, and $P\dsim Q$,
  then there exists a bijection $f$ from $I$ to $J$ such that $\forall
  i\in I.\, S_i\dsim S'_{f(i)}$.  
\end{lem}
\begin{proof}
  We first observe a general property of distributed transitions: for
  any $i_0\in I$, whenever $S_{i_0}\xr{\mu}P_0$, by
  Def.~\ref{def:dsim}, we have $S'_{j_0}\xr{\mu}Q_0$ for some $j_0,
  Q_0$, with $P_0\dsim Q_0$ and $\prod_{i\in I, i\neq
    i_0}S_i\dsim\prod_{j\in J,j\neq j_0}S'_j$, where the latter
  equivalence involves processes that have exactly one parallel
  component less than $P$ and $Q$ respectively. The symmetrical
  property also holds for challenges coming from $Q$.

\medskip
  
Let us now prove that $I$ and $J$ have the same cardinal. We assume
without loss of generality that $I$ has strictly more elements than
$J$. We derive a contradiction by repeatedly using the remark above to
fire challenges in the parallel components of $P$, until there are no
components left in $Q$. $I$ and $J$ thus have the same cardinal.

\medskip

In light of this result, we can assume w.l.o.g.\ that $I$ is the set
of indices in $P$'s and $Q$'s decompositions, and moreover that $0\in
I$. We thus show:
\begin{quotation}
  If $P=\prod_{i\in I}S_i$, $Q=\prod_{j\in I}S'_j$, and $P\dsim Q$,
  then there exists a bijection $f$ from $I$ to $J$ such that $\forall
  i\in I.\, S_i\dsim S'_{f(i)}$.
\end{quotation}
  
To prove this, we reason by induction on the number of parallel
components of $P$.
The cases where this number is $0$ or $1$ are immediate. Assume then
that $I$ has at least two elements. We distinguish two cases:

\noindent\textsl{First case:} all components are equivalent to
each other on each side, that is, $\forall i\in I.\, S_i\dsim S_0$, and
$\forall j\in I.\, S'_j\dsim S'_0$. It remains to show that one of the
$S_i$s is equivalent to one of $S'_j$s: for this, we use the remark
above about distributed transitions to fire all components of $P$ but
one, which gives us that the remaining component is bisimilar to a
component of $Q$.

\noindent\textsl{Second case:} if we define $C=\set{i.\, S_i\dsim
  S_0}$, we have $\emptyset\nsubseteq C\nsubseteq I$ (since otherwise,
we would be in the first case). Define $C'=I\setminus C$, and perform
a sequence of \dsim-challenges on the side of $P$ in order to fire all
components corresponding to $C'$: we are left with $\prod_{i\in C}S_i
\dsim \prod_{i\in D}S'_i$ for some $D\nsubseteq I$.

Since $C\nsubseteq I$, we can apply induction to derive that the
$S_i$s are one to one equivalent to the $S'_j$s, which yields that all
processes in $\set{S_i, i\in C}\cup\set{S'_j, j\in D}$ belong to the
same equivalence class for \dsim{} (and are hence all equivalent to
$S_0$).

Similarly, by firing all components in $C$, we obtain $\prod_{i\in
  C'}S_i\dsim \prod_{i\in D'}S'_i$ for $D'\nsubseteq I$. Again, as
$C'\nsubseteq I$, we have by induction that every element in \set{S_i,
  i\in C'} is in one to one correspondence with an element of
\set{S'_j, j\in D'}. This implies, by definition of $C'$, that none of
the $S'_j$s for $j\in D'$ is equivalent to $S_0$. Hence, we have that
$D\cap D'=\emptyset$, and $D\cup D'= I$ by a cardinality argument; the
announced property follows.
%
\end{proof}

Lemma~\ref{lem:separation} formalises the fact that in absence of
restriction, distributed bisimilarity is discriminating enough to
analyse the maximum degree of parallelism in processes (in particular,
the expansion law is not valid for location sensitive equivalences).

\begin{prop}
  In \microccsp, $P\dsim Q$ if and only if $P\equiv Q$.
\end{prop}
\begin{proof}
  We first remark that $\mathord\equiv\subseteq\mathord\dsim$ on
  \microccsp. To show the converse, we assume $P\dsim Q$, and reason
  by induction on the size of $P$, defined as the number of prefixes
  in $P$. The cases where $P$ is of size $0$ or $1$ are immediate.
  Assume then that the size of $P$ is strictly greater than $1$.
  
  First, if $P$ has at least two parallel components that are
  different from \nil, we can apply the separation property
  (Lemma~\ref{lem:separation}), together with the induction
  hypothesis, to deduce the expected result.

  \medskip
  
  Assume now $P=\sum_{i\in I}\eta_i.P_i$. By
  Lemma~\ref{lem:separation}, $Q$ has only one parallel component,
  i.e., $Q=\sum_{j\in J}\eta'_i.Q_i$. 
  Using the idempotence law $(S+S\equiv S)$, we moreover assume w.l.o.g.
  that for all $i_1, i_2$, $\eta_{i_1}.P_{i_1}\equiv
  \eta_{i_2}.P_{i_2}$ implies $i_1=i_2$, and similarly for the
  summands of $Q$.
  
  Since $P\dsim Q$, we observe two properties. First, $\forall i\in
  I.\,\exists j\in J.\,\eta_i=\eta'_j\land P_i\dsim Q_j$: this follows
  by firing a challenge on $\eta_i$ on $P$'s side. Symmetrically,
  $\forall j\in J.\,\exists i\in I\,\eta_i=\eta'_j\land P_i\dsim Q_j$.
  In each case, the induction hypothesis actually gives $P_i\equiv
  Q_j$.
  
  Now, for any $i_1\in I$, the first property associates some $j\in J$
  to $i$, which in turn is associated to $i_2\in I$ by the second
  property. In this case, we have $\eta_{i_1}=\eta_j=\eta_{i_2}$ and
  $P_{i_1}\equiv Q_j\equiv P_{i_2}$, which insures $i_1=i_2$ by the
  hypothesis we have made. A similar argument, starting from $Q$'s
  side, shows that these two properties entail that the summands in
  $P$ are in one to one correspondence with the summands of $Q$,
  whence, finally, $P\equiv Q$.
%
\iflong
We first establish the following \emph{separation property}:
\begin{quotation}
  If $P=\prod_{i\in I}S_i$, $Q=\prod_{j\in J}S'_j$, and $P\dsim Q$,
  where none of the $S_i, S'_j$ is equivalent to \nil, then there
  exists a bijection $f$ from $I$ to $J$ such that $\forall i\in I.\,
  S_i\dsim S'_{f(i)}$.
\end{quotation}
To show this, we observe that whenever $S_{i_0}\xr{\mu}P_0$, by
Def.~\ref{def:dsim}, we have $S'_{j_0}\xr{\mu}Q_0$ for some $j_0,
Q_0$, with $P_0\dsim Q_0$ and $\prod_{i\in I, i\neq
  i_0}S_i\dsim\prod_{j\in J,j\neq j_0}S'_j$, where the latter
equivalence involves processes that have exactly one parallel
component less than $P$ and $Q$ respectively. 
\damien{je n'ai pas encore vérifié cette preuve} 
Consider now $i_0\in I$, and define $C=\set{i.\, S_i\dsim S_{i_0}}$.
We repeatedly use the observation above by firing all components of
$P$ that are not equivalent to $S_{i_0}$. This yields $\prod_{i\in
  C}S_i\dsim\prod_{j\in D}S'_j$, for some $D\subseteq J$.
Necessarily, $C$ and $D$ have the same cardinality, since otherwise we
could derive a contradiction by playing distributed bisimilarity
challenges on the side where there are strictly more components.
Similarly, if $S_{i_0}\not\dsim S'_j$ for some $j\in D$, we can
isolate this $S'_j$ component (by firing all other components in
$\prod_{j\in D}S'_j$) and derive a contradiction.

Hence for any $i_0$, there are as many parallel components that are
bisimilar to $S_{i_0}$ in $P$ as in $Q$. The separation property
announced above thus follows.

\medskip

We now prove that if $P=\sum_{i\in I}\eta_i.P_i$ and $Q=\sum_{j\in
  J}\eta'_j.Q_j$, then $P\dsim Q$ implies $P\equiv Q$. We reason by
induction on the size of $P$, defined as the number of prefixes in
$P$. Consider $P\xr{\mu}P'$, where $\mu$ is a visible action;
necessarily $\mu=\eta_i$ and $P'=P_i$ for some $i$. Since $P\dsim Q$,
$Q\xr{\eta'_j}Q_j$ for some $j$, with $P_i\dsim Q_j$. By induction,
the latter equivalence gives $P_i\equiv Q_j$. We hence have $\forall
i\in I.\, \exists j\in J.\, (\eta_i=\eta'_j)\land(P_i\equiv Q_j)$, and
symmetrically for all $j\in J$. This is sufficient to conclude
$P\equiv Q$ (we can normalise processes w.r.t.\ the laws $S+0\equiv S$
and $S+S\equiv S$, so that $I$ and $J$ are in bijection as in the
separation property).

\medskip

Hence, if $P\dsim Q$, by erasing topmost parallel components that are
equal to \nil{} in $P$ and $Q$, we can apply the separation property,
and then the result we have just established, to deduce $P\equiv Q$.
Since conversely $\mathord\equiv\subseteq\mathord\dsim$ on \microccsp, we can
conclude.
%
\fi
\end{proof}
  




In view of this result, \dsim{} is arguably not very interesting in
\microccsp. The main point here is to show a situation where $\sim$ is
not substitution closed, while \dsim{} is. It can be proved (but this
requires more work) that the same holds if we move to a richer
calculus, where parallel compositions are allowed in summands. In such
a calculus, \dsim{} satisfies nontrivial \emph{absorption laws}, such
as $a.b\,|\,\out{a}.c~\dsim~ (a.b\,|\,\out{a}.c)+\tau.(b|c)$, which is
obviously not valid for $\equiv$ (we suppose here the existence of a
$\tau$ prefix; more general absorption laws
can be defined -- see~\cite{castellani:phd}).
%

One way to establish that \dsim{} is closed under substitution in the
richer calculus is to exploit the results
of~\cite[Sect.~4.5]{castellani:phd}, which studies axiomatisations
of \dsim. These axiomatisations use a new operator, noted \semipar,
that satisfies the following laws:
\begin{mathpar}
  (P+Q)\semipar R~=~ P\semipar R + Q\semipar R
  \and
  (P\semipar Q)\semipar R ~=~ P\semipar (Q | R)
  \\
  P\semipar\nil ~=~ P
  \and
  \nil\semipar P ~=~\nil
\end{mathpar}
\semipar{} is a kind of \emph{asymmetric parallel composition}, that
intuitively gives precedence to the transitions of its left hand side
operand. Moreover, as shown in~\cite{castellani:phd}, if we allow
communications across \semipar, then the following \emph{expansion
  theorem}
\begin{mathpar}
  ~\quad\mbox{If } P=\sum_{i\in I}\eta_i.P_i\semipar P'_i
  \mbox{ and } Q=\sum_{j\in J}\eta'_j.Q_j\semipar Q'_j
  \mbox{, then}
  \hfill
  \\
  P\,|\,Q\quad =\quad \sum_{i\in I}\eta_i.P_i\semipar (P'_i|Q)
  + \sum_{j\in J}\eta'_j.Q_j\semipar (P|Q'_j)
  + \sum_{\eta_i=\eta'_j}\tau.(P_i|Q_j)\,\semipar\,(P'_i|Q'_j)
\end{mathpar}
\noindent together with the laws of \semipar{} and of $+$, provides a
complete axiomatisation of \dsim.

This expansion theorem closely resembles its standard counterpart in
interleaving semantics, where concurrency is expressed using the sum
operator. However, since communications are allowed between the
operands of \semipar, the above equality is robust w.r.t.\ 
substitution.
Indeed, if a new interaction is triggered on the left hand side of the
equality by applying a substitution, say between $\eta_i.P_i$ and
$\eta'_j.Q_j$, then this synchronisation is also possible on the
right hand side (in the first summand).
We do not enter any further into the details of this proof.



%% file: pi.tex
\subsection{The Finite, Sum-free \texorpdfstring{$\pi$}{pi}-calculus}

Processes of \PI{} are built from an infinite set $\names_\pi$ of
names (we let $a, b\dots, m, n\dots, p, q\dots, x, y\dots$ range over
names), according to the following grammar:
\begin{align*}
  \prefpi &::= m(x) \OR \out{m}{n}\enspace, &%
  P &::= \nil \OR \prefpi.P \OR P_1\|P_2 \OR (\new p)P\enspace.
\end{align*}
The input prefix $m(x)$ binds name $x$ in the continuation process,
and so does name restriction $(\new n)$ in the restricted process. A
name that is not bound is said to be free, and we let \fn{P} stand for
the free names of $P$.  We assume that any process that we manipulate
satisfies a \emph{Barendregt convention}: every bound name is distinct
from the other bound and free names of the process. We shall use $a,
b, c$ to range over free names of processes, $p, q, r$ (resp.\ $x, y$)
to range over names bound by restriction (resp.\ by input), and $m, n$
to range over any name, free or bound (note that these naming
conventions are used in the above grammar).
Structural congruence on \PI, written $\equiv$, is the smallest
congruence that is an equivalence relation, contains
$\alpha$-equivalence, and satisfies the following laws:
\begin{mathpar}
  P\|\nil\equiv P \and %
  P\|(Q\|R)\equiv(P\|Q)\|R \and %
  P\|Q\equiv Q\|P \and %
  (\new p)\nil\equiv\nil \and %
  (\new p)(\new q)P\equiv(\new q)(\new p)P \and %
  P\|(\new p)Q\equiv(\new p)(P\|Q)\quad\text{if}~p\notin\fn{P} %
\end{mathpar}
We let $P[n/x]$ stand for the capture avoiding substitution of name
$x$ with name $n$ in $P$. We use $\sigma$ to range over substitutions
in \PI{} (that simultaneously replace several names).

\begin{defi}[Late operational semantics and
  ground bisimilarity]\label{def:sospinew} The late operational
  semantics of \PI{} is given by a transition relation whose set of
  labels is defined by:
  \begin{align*}
    \mu&::= a(x) \OR \out{a}{b} \OR \bout{a}{p} \OR \tau\enspace.
  \end{align*}
  Names $x$ and $p$ are said to be bound in actions $a(x)$ and
  $\bout{a}{p}$ respectively, and other names are free. We use
  $\bn{\mu}$ (resp.\ \fn{\mu}) to denote the set
  of bound (resp.\ free) names of action $\mu$.
  
  The late transition relation, written $\xlate{}$, is given by the
  following rules (symmetrical versions of the rules involving
  parallel composition are omitted):
  \begin{mathpar}
    \inferrule*{~}{\prefpi.P\xlate{\prefpi}P} \and %
    \inferrule*{P\xlate{a(x)}P'\and Q\xlate{\out{a}{b}}Q'}
    {P\|Q\xlate{\tau}P'[b/x]\|Q'} \\ %
    \inferrule*[right=$a\neq b$]{P\xlate{\out{a}{b}}P'} %
    {(\new b)P\xlate{\bout{a}{b}}P'} \and %
    \inferrule*{P\xlate{a(x)}P'\and Q\xlate{\bout{a}{p}}Q'}
    {P\|Q\xlate{\tau}(\new p)(P'[p/x]\|Q')} \\ %
    \inferrule*[right=$\bn{\mu}\cap\fn{Q}{=}\emptyset$]
    {P\xlate{\mu}P'}{P\|Q\xlate{\mu}P'\|Q} \and %
    \inferrule*[right=$p\notin\fn{\mu}$]%
    {P\xlate{\mu}P'}{(\new p)P\xlate{\mu}(\new p)P'}
  \end{mathpar}
  
  A \emph{ground bisimulation} is a symmetric relation $\RR$ between
  processes such that whenever $P\RR Q$ and $P\xlate\mu P'$, there
  exists $Q'$ s.t.\ $Q\xlate\mu Q'$ and $P'\RR Q'$.
  
  \emph{Ground bisimilarity}, written \simgr, is the union of all
  ground bisimulations.
\end{defi}
Note that we do not respect the convention on names in the rule to
infer a bound output, precisely because we are transforming a free
name ($b$) into a bound name.

\begin{lem}\label{lem:trans:sub}
  Assume that $P\sigma\xlate{\mu}P'$.
  \begin{enumerate}[\em(1)]
  \item\label{sub:one} If $\mu$ is $\out{a}{b}$, \bout{a}{p} or
    $a(x)$, then $P\xlate{\mu'}P''$ with $\mu'\sigma=\mu$ and
    $P''\sigma=P'$.
  \item\label{sub:two} If $\mu=\tau$ then one of the three following
    properties hold, where the input and output actions are offered
    concurrently by $P$ in the last two cases. 
    \medskip
    \begin{enumerate}[\em(a)]
    \item\label{sub:two:one} $P\xlate{\tau}P''$ and $P''\sigma=P'$,
    \item\label{sub:two:two} $P\xlate{\out{b}{c}}\xlate{a(x)}P''$ where
      $\sigma(a)=\sigma(b)$ and $P''[c/x]\sigma\sim P'$,
    \item\label{sub:two:three} $P\xlate{\bout{b}{p}}\xlate{a(x)}P''$ where
      $\sigma(a)=\sigma(b)$ and $((\new p)P''[p/x])\sigma\sim P'$.
    \end{enumerate}
  \end{enumerate}
\end{lem}
\begin{proof}
  Similar to the proof of Lemma 1.4.13 in~\cite{SW01}, where the early
  transition semantics is treated. 
\end{proof}

\subsection{\texorpdfstring{\majMD}{majMD}{}s in \texorpdfstring{\PI}{PI}}

In what follows, we fix two distinct names $a$ and $b$, that will
occur free in the processes we shall consider. The definitions and
results below will depend on $a$ and $b$, but we avoid making this
dependency explicit, in order to ease readability.
Names $a$ and $b$ will be fixed in the proof of
Theorem~\ref{thm:closure:simg}.

\begin{defi}[Erasing a \PI{} process]
  Given a \PI{} process $P$, we define the \emph{erasing of $P$},
  written \erase{P}, as follows:
  \begin{align*}
    \erase{P_1\|P_2} &\eqdef \erase{P_1}\|\erase{P_2} &
    \erase{(\new p)P} &\eqdef \erase P &
    \erase{\nil} &\eqdef \nil \\
    \erase{a(x).P} &\eqdef a.\erase P &  
    \erase{m(x).P} &\eqdef \nil~\text{if}~m\neq a \\
    \erase{\out{b}{n}.P} &\eqdef \overline{b}.\erase P &
    \erase{\out{m}{n}.P} &\eqdef \nil~\text{if}~m\neq b 
  \end{align*}
\end{defi}

Note that $a$ and $b$ play different roles in the definition of
\erase{\cdot}.

It is immediate from the definition that \erase{P} is a \microccs{}
process whose only prefixes are $a$ and $\overline{b}$.
Intuitively, \erase{P} only exhibits the interactions of $P$ at $a$
(in input) and $b$ (in output) that are not guarded by interactions on
other names.

\begin{lem}[Transitions of \erase{P}]\label{lem:trans:erase}
  Consider a \PI{} process $P$. We have: 
  \begin{enumerate}[$\bullet$]
  \item If $P\xlate{a(x)}P'$, then $\erase{P}\xr{a}\erase{P'}$.
  \item If $P\xlate{\out{b}{c}}P'$ or $P\xlate{\bout{b}{p}}P'$, then
    $\erase{P}\xr{\overline{b}}\erase{P'}$.
  \item Conversely, if $\erase{P}\xr{a}P_0$, then there exist $x$ and
    $P'$ such that $P_0=\erase{P'}$ and $P\xlate{a(x)}P'$. Similarly, if
    $\erase{P}\xr{\overline{b}}P_0$, there exist $c, p, P'$ such that 
    $P_0=\erase{P'}$ and either $P\xlate{\out{b}{c}}P'$ or
    $P\xlate{\bout{b}{p}}P'$.
  \end{enumerate}
\end{lem}
\begin{proof}
  Simple reasoning on the LTSs of \microccs{} and \PI. 
\end{proof}

\begin{prop}[Transfer]\label{prop:transfer2ccs}
  If $P\simg Q$ in \PI, then $\erase{P}\sim\erase{Q}$ in \microccs.
\end{prop}
\begin{proof}
  We reason by induction on the size of $P$ (defined as the number of
  prefixes in $P$). Consider a transition of \erase{P}; as observed
  above, it can only be a transition along $a$ or a transition along
  $\overline{b}$.

  Assume $\erase{P}\xr{a}P_0$. By Lemma~\ref{lem:trans:erase},
  $P\xlate{a(x)}P'$ and $P_0=\erase{P'}$. Since $P\simg Q$,
  $Q\xlate{a(x)}Q'$ for some $Q'$ such that $P'\simg Q'$. By induction, the
  latter relation gives $\erase{P'}\sim\erase{Q'}$, and
  $Q\xlate{a(x)}Q'$ gives by Lemma~\ref{lem:trans:erase}
  $\erase{Q}\xr{a}\erase{Q'}$.
  
  The case $\erase{P}\xr{\overline{b}}P_0$ is treated similarly: by
  Lemma~\ref{lem:trans:erase}, there are two cases, according to
  whether $P$ does a free output or a bound output. Reasoning like
  above allows us to conclude in both cases. 
\end{proof}

We can now present our central technical result about \PI, which comes
in two lemmas.

\begin{lem}\label{lem:grange:pi:external}
  If $Q\simg(\new\ps)(a(x).P_1\|\out{b}{c}.P_2\|P_3)$, then there
  exist some $Q_1$, $Q_2$, $Q_3$, $\qs$, such that
  $Q\equiv (\new \qs)(a(x).Q_1\|\out{b}{c}.Q_2\|Q_3)$ and 
  \begin{align*}
    (\new\ps)(P_1\|P_2\|P_3)~\simg~(\new\qs)(Q_1\|Q_2\|Q_3).
  \end{align*}
\end{lem}
\begin{proof}
  Let $P=(\new\ps)(a(x).P_1\|\out{b}{c}.P_2\|P_3)$ and
  $P'=(\new\ps)(P_1\|P_2\|P_3)$.

  Note that by our conventions on notations, $c\notin\ps$.
  
  Since $Q\simg P$ and $P$ can perform two transitions along $a(x)$
  and $\out{b}{c}$ respectively, $Q$ can also perform these
  transitions, which gives

  $Q\equiv(\new\qs)(a(x).Q_1\|\out{b}{c}.Q_2\|Q_3)$ for some
  $\qs,Q_1,Q_2,Q_3$,

  \noindent the first (resp.\ second) component exhibiting the
  prefix that is triggered to answer the challenge on $a(x)$ (resp.
  \out{b}{c}).
  
  Consider now the challenge $P\xlate{\out{b}{c}}\xlate{a(x)}P'$, to
  which $Q$ answers by performing the transition
  $Q\xlate{\out{b}{c}}\xlate{a(x)}Q_{ba}$, with $P'\simg Q_{ba}$. If
  $Q_{ba} = (\new\qs)(Q_1\|Q_2\|Q_3)$, that is, if $Q$ triggers the
  prefixes on top of its first and second components, then we are
  done. Similarly, if $Q$ triggers a prefix in $Q_3$ to answer the
  second challenge, say $Q_3 = a(x).Q_4\|Q_5$, we can set $Q'_1 =
  a(x).Q_4$ and $Q'_3 = Q_1\|Q_5$, and the lemma is proved.
  
  The case that remains to be analysed is when $Q_2\xlate{a(x)}Q'_2$
  and we have $Q_{ba} =
  (\new\qs)(a(x).Q_1\|Q'_2\|Q_3)\simg(\new\ps)(P_1\|P_2\|P_3)$.

  \medskip
  
  We then consider the challenge where $P$ fires its two topmost
  prefixes $a(x)$ and $\out{b}{c}$ in the other sequence, namely
  $P\xlate{a(x)}\xlate{\out{b}{c}}P'$. By hypothesis, $Q$ triggers the
  prefix of its first component for the first transition. To perform
  the second transition, $Q$ can fire the prefix $\out{b}{c}$ either
  in its second or third component, in which case, as above, we are
  done, or, and this is the last possibility, the prefix $\out{b}{c}$
  occurs in $Q_1$. This means $Q_{ab} =
  (\new\qs)(Q'_1\|\out{b}{c}.Q_2\|Q_3)\simg(\new\ps)(P_1\|P_2\|P_3)$,
  with $Q_1\xlate{\out{b}{c}}Q'_1$.

  \medskip
  
  To sum up, we have $Q_{ab} =
  (\new\qs)(Q'_1\|\out{b}{c}.Q_2\|Q_3)\simg
  (\new\qs)(a(x).Q_1\|Q'_2\|Q_3) = Q_{ba}$, with
  $Q_1\xlate{\out{b}{c}}Q'_1$ and $Q_2\xlate{a(x)}Q'_2$: this
  resembles the \MD{} of Definition~\ref{def:MD}, translated into the
  $\pi$-calculus. 
  
  Indeed, we can construct a \MD{} in \microccs: $Q_{ab}\simg Q_{ba}$
  implies $\erase{Q_{ab}}\sim\erase{Q_{ba}}$ by
  Prop.~\ref{prop:transfer2ccs}, and $Q_1\xlate{\out{b}{c}}Q'_1$
  (resp. $Q_2\xlate{a(x)}Q'_2$) implies by Lemma~\ref{lem:trans:erase}
  $\erase{Q_1}\xr{\overline{b}}\erase{Q'_1}$ (resp.
  $\erase{Q_2}\xr{a}\erase{Q'_2}$). Finally, using
  Lemma~\ref{lem:grange}, we obtain a contradiction, which concludes
  our proof. 
\end{proof}

\begin{lem}\label{lem:grange:pi:internal}
  If $Q\simg(\new p,\ps)(a(x).P_1\|\out{b}{p}.P_2\|P_3)$, then there
  exist some $Q_1$, $Q_2$, $Q_3$, such that $Q\equiv (\new
  p,\qs)(a(x).Q_1\|\out{b}{p}.Q_2\|Q_3)$ and
  \begin{align*}
    (\new\ps)(P_1\|P_2\|P_3)~\simg~(\new\qs)(Q_1\|Q_2\|Q_3).
  \end{align*}
\end{lem}
\begin{proof}[Hint]
  The proof follows the same lines as for the previous lemma. The only
  difference is when analysing the transitions that lead to $Q_{ab}$:
  to perform the second transition, $Q$ can either extrude the name
  called $p$ in the equality $Q\equiv(\new
  p,\qs)(a(x).Q_1\|\out{b}{p}.Q_2\|Q_3)$, or otherwise $Q$ can be
  $\alpha$-converted in order to extrude another name. In the case
  where $Q$ chooses to extrude a different name, we can assume
  without loss of generality that the necessary $\alpha$-conversion is
  a swapping between name $p$ and a name $q_1\in\qs$, which brings us
  back to the case where name $p$ is the one being extruded.
  
  The presence of a bound output introduces some notational
  complications when expressing $Q_{ab}$, but basically it does not
  affect the proof w.r.t.\ the proof of
  Lemma~\ref{lem:grange:pi:external}, because the function
  \erase{\cdot} is not sensitive to name permutations that do not
  involve $a$ or $b$. 
\end{proof}

\subsection{Congruence}

\begin{thm}[Closure of \simg{} under
  substitution]\label{thm:closure:simg} 
  If $P\simg Q$ then for any substitution $\sigma$, $P\sigma\simg
  Q\sigma$.
\end{thm}
\begin{proof}
  We prove that the relation $\s{\RR}\eqdef\{(P\sigma,
  Q\sigma) \mid P\simg Q\}$ is a ground bisimulation. We consider $P$,
  $Q$ such that $P\simg Q$ and assume $P\sigma\xlate{\mu}P_0$.
  We examine the transitions of $P$ that make it possible for
  $P\sigma$ to do a $\mu$-transition to $P_0$.

  \medskip

  According to Lemma~\ref{lem:trans:sub}, there are two possibilities.
  The first possibility corresponds to the situation where $\mu$ comes
  from an action that $P$ can perform, i.e., $P\xlate{\mu'}P'$ for
  some $\mu'$, with $P'\sigma=P_0$ and $\mu'\sigma=\mu$
  (cases~\ref{sub:one} and~\ref{sub:two:one} in
  Lemma~\ref{lem:trans:sub}).  Since $P\simg Q$, $Q\xlate{\mu'} Q'$
  and $P'\simg Q'$ for some $Q'$. We can prove that $Q\sigma\xr\mu
  Q'\sigma$, and since $P'\simg Q'$ we have $(P'\sigma,
  Q'\sigma)\in\s{\RR}$.

  \smallskip

  The second possibility (which corresponds to the difficult case) is
  given by $\mu=\tau$, where the synchronisation in $P'$ has been made
  possible by the application of $\sigma$.
  There are in turn two cases, corresponding to whether the
  synchronisation involves a free or a bound name. In the former case,
  $P\xlate{a(x)}P'$ and $P\xlate{\out{b}{c}}P''$ for some $a, x, b, c,
  P', P''$. This entails
  $P\equiv(\new\ps)(a(x).P_1\|\out{b}{c}.P_2\|P_3)$ for some $\ps, P_1,
  P_2, P_3$, and, since $P\simg Q$, we conclude by
  Lemma~\ref{lem:grange:pi:external} that $Q\equiv (\new
  \qs)(a(x).Q_1\|\out{b}{c}.Q_2\|Q_3)$ and
  $$
  (\new\ps)(P_1\|P_2\|P_3)~\simg~(\new\qs)(Q_1\|Q_2\|Q_3)\enspace.
  $$
  By definition of $\RR$, this equivalence implies that we can apply
  any substitution to these two processes to yield processes related
  by $\RR$, and in particular $[c/x]\sigma$, which gives:
  $$
  ((\new\ps)(P_1\|P_2\|P_3))[c/x]\sigma~%
  \RR~((\new\qs)(Q_1\|Q_2\|Q_3))[c/x]\sigma\enspace.
  $$
  Using the Barendregt convention hypothesis, this amounts to
  $$
  P_0\equiv((\new\ps)(P_1[c/x]\|P_2\|P_3))\sigma~%
  \RR~((\new\qs)(Q_1[c/x]\|Q_2\|Q_3))\sigma\eqdef Q_0\enspace.
  $$ 
  We can then conclude by checking that $Q\sigma\xlate\tau Q_0$.

  \medskip

  We reason similarly for the case where the synchronisation involves
  the transmission of a bound name, using
  Lemma~\ref{lem:grange:pi:internal} instead of
  Lemma~\ref{lem:grange:pi:external}. We remark that
  Lemma~\ref{lem:grange:pi:internal} gives
  $(\new\ps)(P_1\|P_2\|P_3)\simg(\new\qs)(Q_1\|Q_2\|Q_3)$, and in this
  case $P\sigma\xlate{\tau}(\new p,\ps)(P_1[p/x]\|P_2\|P_3)\sigma$
  (resp.  $Q\sigma\xlate{\tau}(\new
  p,\qs)(Q_1[p/x]\|Q_2\|Q_3)\sigma$).  In order to be able to add the
  restriction on $p$ to the terms given by
  Lemma~\ref{lem:grange:pi:internal}, we rely on the fact that $\simg$
  is preserved by restriction: $P\simg Q$ implies $(\new p)P\simg
  (\new p)Q$ for any $P, Q, p$. We can then reason as above to
  conclude. 
\end{proof}

\begin{cor}[Congruence of bisimilarity in \PI]
  In \PI, ground, early and late bisimilarity coincide and are
  congruences.
\end{cor}
\begin{proof}
  By a standard argument (see~\cite{SW01}): since $\simg$ is closed
  under substitution, $\simg$ is an open bisimulation. 
\end{proof}

It is known (see~\cite{SW01}) that adding either replication or sum to
\PI{} yields a calculus where strong bisimilarity fails to be a
congruence.



%% file: concl.tex
We have presented an axiomatisation of strong bisimilarity on a small
subcalculus of CCS, and a new congruence result for the
$\pi$-calculus. 

\medskip

Technically, the notion of \MD{} is related to substitution closure of
strong bisimilarity, as soon as substitutions can create new
interactions by identifying two names. 

We have shown in
Sect.~\ref{sec:pi} that there exists no \MD{} in \PI, and that
\simg{} is a congruence.
%
%
%
It appears that in finite calculi, \MD{}s give rise to counterexamples
to substitution closure of strong bisimilarity (cf.\
Sect.~\ref{sec:MDs}). The situation is less clear when infinite
behaviours can be expressed.
{For instance, in the extension of \microccs{} with replication,
  the process $P\eqdef\;!a\|!\overline{b}$ is bisimilar to process
  $Q\eqdef\;!a.\overline{b}\|!\overline{b}.a$, which leads to a \MD:
  we have $Q~\s{\xr{a}\xr{\overline{b}}}\equiv
  Q~\s{\xr{\overline{b}}\xr{a}}\equiv Q$. This \MD{} is however
  `benign': by firing concurrently the two prefixes that iniate the
  \MD{}, we obtain $a\|\out b\|P$ which is bisimilar to $P$, so that
  this situation is not problematic w.r.t.\ substitution closure (we
  may moreover remark that the two aforementioned processes remain
  bisimilar when $b$ is replaced
  with $a$).} %
We do not know at present whether $\sim$ is substitution-closed in
this extension of \microccs{}.

Some subcalculi of the $\pi$-calculus where strong bisimilarity is a
congruence are obtained by restricting the output prefix~\cite{SW01}.
In the \emph{asynchronous $\pi$-calculus} ($A\pi$), 
\MD{}s do not appear, basically because the output action is not a
prefix.  Strong bisimilarity is a congruence on $A\pi$.
In the \emph{private $\pi$-calculus} ($P\pi$), since only private
names are emitted, no substitution generated by a synchronisation can
identify two previously distinct names. Hence, although \MD{}s exist
in $P\pi$ (due to the presence of the sum operator), strong
bisimilarity is not substitution closed, but is a congruence. Indeed,
to obtain the latter property, we only need to consider the particular
substitutions at work in $P\pi$, which cannot identify two names.


\medskip

Regarding future extensions of this work, we would like to study
whether our approach can be adapted to analyse weak bisimilarity in
\PI{} (as mentioned in Remark~\ref{rk:weak}, strong and weak
bisimilarity coincide in \microccs). Another interesting direction, as
hinted above, would be to study strong bisimilarity on infinite,
restriction-free calculi (in CCS and the $\pi$-calculus).
